\documentclass[journal]{vgtc}                     
\title{POEM: Interactive Prompt Optimization for Enhancing Multimodal Reasoning of Large Language Models}

\author{
  Jianben He, Xingbo Wang, Shiyi Liu, Guande Wu, Claudio Silva, and Huamin Qu
}

\authorfooter{

}

\abstract{
Large language models (LLMs) have exhibited impressive abilities for multimodal content comprehension and reasoning with proper prompting in zero- or few-shot settings.
Despite the proliferation of interactive systems developed to support prompt engineering for LLMs across various tasks, most have primarily focused on textual or visual inputs, thus neglecting the complex interplay between modalities within multimodal inputs.
This oversight hinders the development of effective prompts that guide models' multimodal reasoning processes by fully exploiting the rich context provided by multiple modalities.
In this paper, we present \name, a visual analytics system to facilitate efficient prompt engineering for \jianben{steering} the multimodal reasoning performance of LLMs. 
The system enables users to explore the interaction patterns across modalities at varying levels of detail for a comprehensive understanding of the multimodal knowledge elicited by various prompts. 
Through diverse recommendations of demonstration examples and instructional principles, \name supports users in iteratively crafting and refining prompts to better align and enhance model knowledge with human insights.
The effectiveness and efficiency of our system are validated through two case studies and interviews with experts.
 
}


\teaser{
  \centering
  \includegraphics[width=\linewidth]{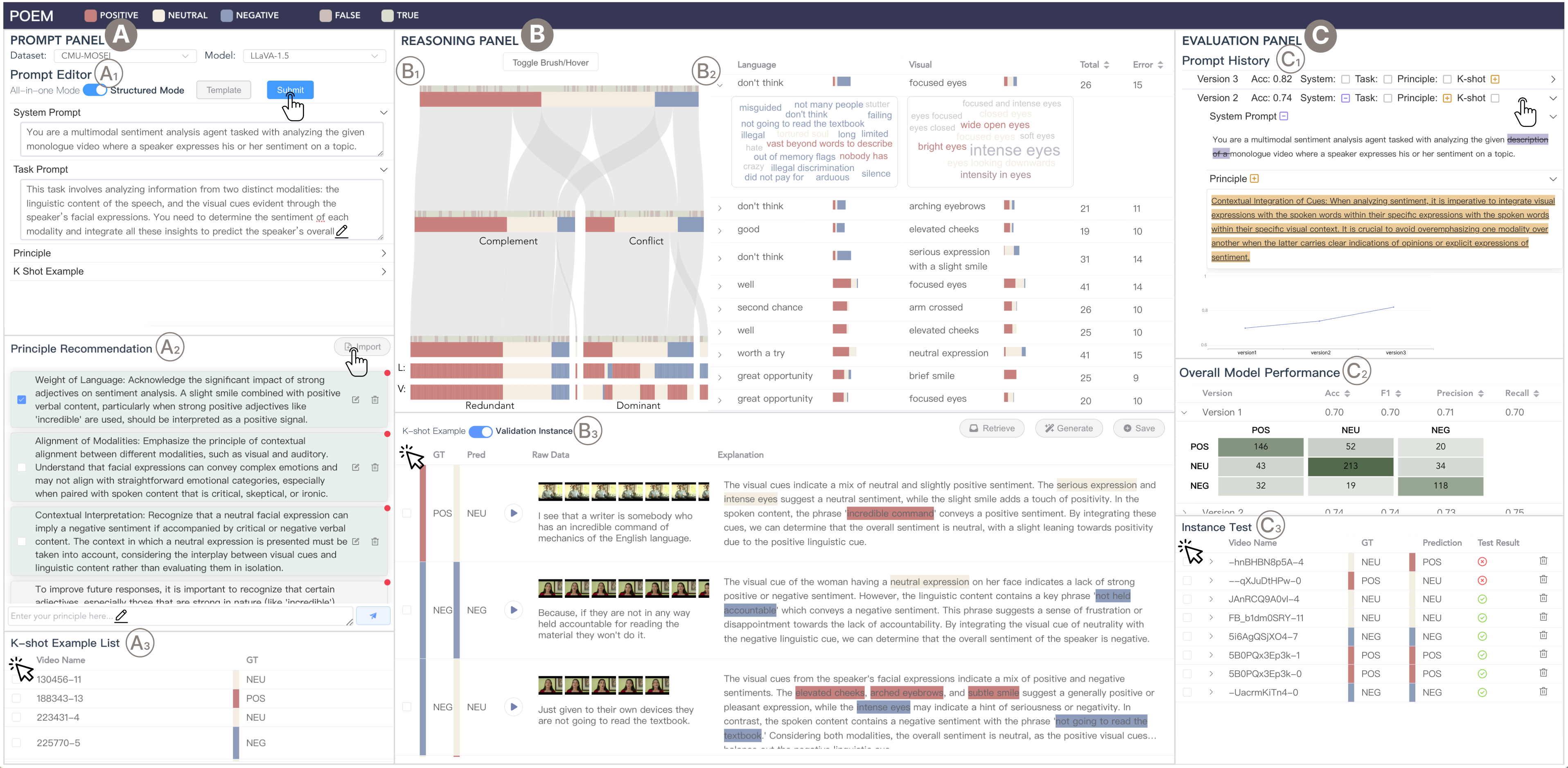}
  \caption{
The \name interface consists of three major panels. The \textit{Prompt Panel} (A) offers versatile operations for users to efficiently craft and edit prompt content, such as importing various principles and demonstration examples, to support an effortless prompt engineering experience.
The \textit{Reasoning Panel} (B) facilitates a comprehensive multi-level investigation of the model's multimodal reasoning performance, ranging from the global modality interaction level to the local instance level.
The \textit{Evaluation Panel} (C) supports both global and local evaluation of prompts, coupled with detailed documentation of modifications during prompt iterations for continuous monitoring and comparison.
}
  \label{fig:teaser}
}

\graphicspath{{figs/}{figures/}{pictures/}{images/}{./}} 
\usepackage[svgnames,dvipsnames]{xcolor}
\usepackage{tabu}                      
\usepackage{booktabs}                  
\usepackage{lipsum}                    
\usepackage{mwe}                       
\usepackage{graphicx}
\usepackage{mathptmx}                  
\usepackage{amsmath}
\usepackage{eucal}

\usepackage{xspace,xpunctuate}
\usepackage{comment}
\usepackage{tabularx,diagbox}
\usepackage{multirow}
\usepackage[noend]{algorithmic}

\newcommand{\xingbo}[1]{{\color{black} #1}}

\newcommand{\jianben}[1]{{\color{black} #1}}

\newcommand{\etal}{\xspace\textit{et al.}\xspace}

\newcommand{\name}{\textit{POEM}\xspace}

\begin{document}



\firstsection{Introduction}
\maketitle
Large Language Models (LLMs), pre-trained on massive data with billions of parameters, have become a cornerstone for natural language processing.
They encode extensive knowledge about the world in their parameter space, exhibiting impressive capabilities in text understanding, reasoning, and generation across various downstream tasks~\cite{tom2020gpt, yang2023foundation}.
Building on the strength of LLMs, there are an increasing number of works~\cite{li2023blip, zeng2023socratic, bhattacharyya2023video, hanu2023language, wang2022language, wang2022language} 
exploring their applications in a wide spectrum of multimodal tasks (e.g., multimodal scene understanding and question answering). 
By using text as the universal representation, 
these works aim to leverage LLMs to integrate and analyze knowledge distilled from diverse modalities (e.g., audio and images) in text format and provide a holistic understanding of multimodal content. 
These models are also known as multimodal LLMs~\cite{yin2023survey, sun2023survey}. 
Comprehending multimodal content necessitates extensive multimodal knowledge, where models not only need to understand the information presented in each individual modality but also have to correctly infer how the information combines to inform accurate reasoning~\cite{yang2023mmbigbench}. 

Recently, prompting has emerged as a data-efficient and user-friendly paradigm for steering and \jianben{improving} LLM's performance on complex reasoning tasks. Relying on extensive knowledge acquired during pre-training, the models can instantly adapt to new downstream tasks in the few-shot or even zero-shot settings without the need for model retraining~\cite{liu2023pretrain}.  
Moreover, the LLMs can be prompted to generate \jianben{free-text rationales} emulating human thought processes in the chain-of-thought (CoT) manner~\cite{bhattacharyya2023video}. 
For example, they can provide step-by-step derivations that lead to the final answer of a math problem or substantiate their analysis process with supported evidence such as emotive words for sentiment prediction~\cite{zhao2024explainability}.
These human-readable rationales enhance the accuracy and transparency of the model's reasoning process~\cite{madsen2022posthoc, mishra2023characterizing}.

While multimodal LLMs exhibit remarkable performance in various tasks with prompting, their reasoning performance is notably sensitive to prompt variations. Moreover, inadequate or ill-designed prompts may elicit erroneous knowledge, resulting in biased and unreliable reasoning. 
Devising well-performing prompts that can guide and \jianben{improve} the multimodal reasoning performance of LLMs remains a persistent challenge. 
Prompting is inherently a process requiring expertise and trial-and-error, where users need to meticulously craft prompts, scrutinize the outputs to identify flaws requiring improvement, and iteratively refine the prompts to reach the intended outcomes~\cite{hendrik2023promptide,dong2022survey}. 
During the process, \textit{users first face the challenge of systematically understanding and examining the multimodal reasoning performance of different prompts. }
Manually inspecting each instance is not only time-consuming but also fails to provide a holistic understanding. Therefore, summarizing and presenting generated rationales at varying detail levels is non-trivial for users to fully verify outputs and pinpoint problematic aspects. However, in the multimodal context, the complex interplay among different modalities, coupled with the unstructured and generative nature of free-text \jianben{rationales}, makes interpreting multimodal LLMs' reasoning process particularly challenging.
Furthermore, \textit{users also struggle to revise their prompts in a way that effectively incorporates and elicits the desired multimodal reasoning knowledge from the model}~\cite{jiang2022promptmaker}. Well-clarified task instructions (e.g., format, phrasing, and content) and informative demonstration examples are both imperative for enabling LLMs to grasp the intended input-output relationships and generate consistent outputs with correct rationales~\cite{mishra2023promptaid}. 
Considering the huge space of possible task instructions and the difficulty of selecting and annotating demonstration examples from high-dimensional and information-complex multimodal data, it is important yet challenging to facilitate users to craft and refine prompts in an efficient manner.

To tackle the above challenges, we present \name, a visual analytics approach designed to streamline the process of prompt engineering for \jianben{model practitioners, including model developers and model users}, to systematically probe and steer the multimodal reasoning performance of LLMs for targeted downstream tasks.
\xingbo{To build a comprehensive understanding of LLMs' knowledge and reasoning on multimodal tasks, we develop computational methods to decompose and summarize cross-modal interactions captured by LLMs in various levels of detail.
At the modality level, we adopt a three-layer augmented Sankey diagram to contextualize model performance with complement and conflict interactions between modalities.
Then, drilling down into specific interactions, we distill and summarize the linguistic and visual evidence from individual instances to reflect model reasoning patterns.
These visualizations help align the model's knowledge and reasoning processes with the human understanding at scale.}
\xingbo{Based on the multi-level model understanding, \name{} allows users to conduct both top-down and bottom-up approaches to build and refine prompts that guide LLM's multimodal reasoning.} 
Specifically, we employ an effective sampling strategy for demonstration examples, ensuring a balance between relevance and diversity to provide varied and informative input-output mappings for inductive model learning. 
On the other side, drawing on human innate capabilities for summarization and generalization, we incorporate an LLM-assisted module that distills principles at both instance-specific and agnostic levels. This approach facilitates users to precisely articulate and apply their domain-specific knowledge and expertise to guide the model deductively.

Our contributions are summarized as follows:
\begin{itemize}[itemsep=0pt,topsep=0pt,parsep=0pt]
\item We propose an effective human-in-the-loop workflow that facilitates systematic investigation and guidance of the multimodal reasoning performance of LLMs.
\item We develop a visual analytics system \name, equipped with carefully designed visualizations and interactions to support efficient prompt engineering for multimodal reasoning tasks.
\item We conduct two case studies and expert interviews to demonstrate the usefulness and efficiency of \name.
\end{itemize}

\section{Related works}
\label{sec:related_works}
The research studies related to the design of \name{} include prompt engineering, multimodal reasoning, and visual analytics for model understanding and steering.
\subsection{Prompt Engineering}
Equipped with extensive knowledge acquired during pre-training, LLMs (e.g., GPT-\cite{tom2020gpt} and LLaMA\cite{touvron2023llama} series models)
exhibit remarkable adaptability to specialized downstream tasks such as question answering, content retrieval, and complex reasoning, given precise instructions and proper demonstration examples. 
This emerging paradigm, known as prompting or prompt engineering, offers a user-friendly and data-efficient way for non-expert users to interact and steer large models.
Prompting generally includes instruction-based and example-based prompts~\cite{bhattacharjya2024foundation}. Instruction-based prompts include system prompts that provide general guidelines and task prompts that deliver direct and task-specific instructions. Example-based prompts utilize a small set of examples to showcase the desired input-output patterns for models to follow.
Numerous studies~\cite{jiang2022promptmaker, jd2023jonny, sherry2023scattershot} have highlighted two major challenges in prompting: the formulation of effective prompts, and the assessment of prompt efficacy alongside strategies for enhancement.


Many studies have been conducted to address the prompting challenges.
Strobelt~\etal introduced PromptIDE~\cite{hendrik2023promptide} as a tool for rapid exploration and assessment of variations in prompt templates.
KnowledgeVis~\cite{coscia2023knowledgevis} compared multiple fill-in-the-blank prompts to probe the input-output associations in BERT-based models. 
Beyond expediting the wording and phrase structure refinement, ScatterShot~\cite{sherry2023scattershot} proposed a slice-based sampling strategy to identify the most informative data patterns for human annotation.
PromptAid~\cite{mishra2023promptaid} combined multiple prompt perturbation strategies to find satisfactory prompts for text classification tasks. 
In addition, PromptChainer~\cite{wu2022promptchainer} and AI Chains~\cite{sherry2022aichains} have been developed to support more sophisticated tasks by decomposing them into manageable sub-tasks, and supported prompt chain prototyping and authoring to enhance controllability.
Kim\etal proposed EvaLM\cite{kim2023evallm} for iterative prompt evaluation according to user-defined criteria, while ContitutionMaker~\cite{petridis2023constitutionmaker} converted users' natural language feedback into a principle for chatbot behavior customization. 
Besides text-to-text generative tasks, several works facilitated prompt refinement for text-to-image generation by keywords~\cite{feng2024promptmagician} and style description recommendation~\cite{brade2023promptify}, structured search of visual concepts~\cite{liu2022opal}, and rubric-based adjustment for precise emotion expression~\cite{wang2023reprompt}. 

However, existing interactive prompt engineering systems are limited to text-to-text or text-to-image generation tasks, failing to deal with the complexity of multimodal inputs for more sophisticated reasoning tasks.
In this paper, we develop \name to optimize the prompt engineering process for adapting and steering the multimodal reasoning performance of LLMs. 
\name facilitates a comprehensive investigation of prompt effects and provides diverse support for users to iterate prompts with reduced cognitive burden and increased efficiency.




\subsection{Multimodal Reasoning}
Reasoning generally refers to the process of drawing on evidence to make logical inferences based on existing knowledge for prediction and decision-making~\cite{sun2023survey}.
In the multimodal context, it is imperative for models to not only grasp evidence derived from single modalities but also to comprehend how evidence from different modalities relates to each other. 
This comprehension could lead to the generation of new insights that the models must capture to achieve accurate reasoning. 
Recently, LLMs have demonstrated the capability to generate coherent \jianben{rationales} through Chain of Thought (CoT) prompting~\cite{wei2022cot}, \jianben{where LLMs provide the intermediate reasoning steps in natural language that lead to the final answer}~\cite{zhao2024explainability}. 
These generated \jianben{free-text rationales} have been increasingly explored for model \jianben{interpretability}, as they provide an explicit and transparent way to communicate the decision-making process of models to end-users in a human-like manner~\cite{mishra2023characterizing}.

A growing number of benchmark datasets~\cite{lu2022scienceqa, yue2023mmmu, fu2024video} have been proposed to evaluate the capabilities of LLMs in multimodal reasoning tasks, with a primary focus on visual content understanding like Visual Question Answering (\jianben{i.e., answering text questions based solely on visual content}). However, the comprehension of how LLMs integrate and coordinate information from various modalities (visual + language, or additional modalities) in the given context for question answering and reasoning remains under-explored. This includes tasks that necessitate a nuanced understanding of multimodal contexts, such as multimodal scene comprehension and multimodal sentiment analysis~\cite{yang2023mmbigbench}.
Moreover, the metrics on these benchmarks fail to capture the detailed reasoning process of models for in-depth model understanding and diagnosis. 
Besides evaluation, many works~\cite{lu2023chameleon, zeng2023socratic, Shao_2023_CVPR, wang2022language} have tried to steer multimodal LLMs' reasoning abilities. 
Compared with the labor-intensive fine-tuning approaches involving curating specific datasets with additional reasoning chain annotation, the training-free prompting-based methods~\cite{zheng2023ddcot, zhang2024context} have become prevalent. 
However, these automatic techniques fall short of providing fine-grained prompt evaluation and flexible prompt refinement. 
Instead, we present a human-in-the-loop approach where users can interactively examine, evaluate, and refine prompts to guide and \jianben{steer} model performance in a more interpretable and controllable manner.

\subsection{Visual Analytics for model understanding and steering}
Visual Analytics has proved to be an effective approach to help users understand and steer machine learning models~\cite{yuan2021survey, yang2023foundation}.
Prior works aimed to disclose the functionalities of neurons and layers of diverse neural network models like RNNs~\cite{ming2017rnnvis, strobelt2018lstmvis, hohman2020summit}. 
\jianben{Recently, many works~\cite{derose2021attentionflow, li2021t3, wang2021dodrio, hoover2020exbert, yeh2023attentionviz, jaunet2022visqa, liu2018nlize} have sought to elucidate the attention mechanism to understand the inner workings of transformer-based models in reasoning and decision-making process.}
Beyond visualizing model internals, numerous studies~\cite{wexler2020whatif, zhang2023sliceteller, wang2022m2lens, liang2023multiviz, cabrera2023zeno,tenney2020lit, li2022unified} have tried to probe model knowledge through analyzing post-hoc model behaviors with input variations.
For example, 
\jianben{M2Lens~\cite{wang2022m2lens} and MultiViz~\cite{liang2023multiviz} characterized intra- and inter-modal interactions with aggregated feature importance for multimodal model diagnosis.}
The What-If Tool~\cite{wexler2020whatif} and SliceTeller~\cite{zhang2023sliceteller} identified specific data slices to understand model failures. 
Integrated tools~\cite{tenney2020lit, cabrera2023zeno, li2022unified} have also been developed for unified language model evaluation. 

Beyond mere understanding, recent works~\cite{boggust2022shared, hoque2023visual, he2024videopro, wang2024commonsense} have progressed  to align model behavior with human knowledge, thereby adapting and steering models to generate desired outcomes for specific tasks. SharedInterest~\cite{boggust2022shared} designed quantitative metrics using saliency methods to compare human and model reasoning for identifying recurring model behavior patterns.
Hoque\etal~\cite{hoque2023visual} and He\etal~\cite{he2024videopro} employed data programming concepts to inject human knowledge at scale for model improvement.
CommonsenseVis~\cite{wang2024commonsense} constructed knowledge graphs with external knowledge bases to contextualize model \jianben{reasoning} behaviors and allow interactive model editing to enhance specific knowledge for poorly behaved areas. 
\jianben{Our work expands on these ideas to examine post-hoc model behaviors with varied prompt inputs for comprehending how different prompts affect model performance. It further enables model practitioners to provide feedback and align model performance with their knowledge and expertise through iterative prompting.}

\section{Design Requirements}
\label{sec: design requirements}
\begin{figure*}[ht]
    \centering
    \includegraphics[width=\linewidth]{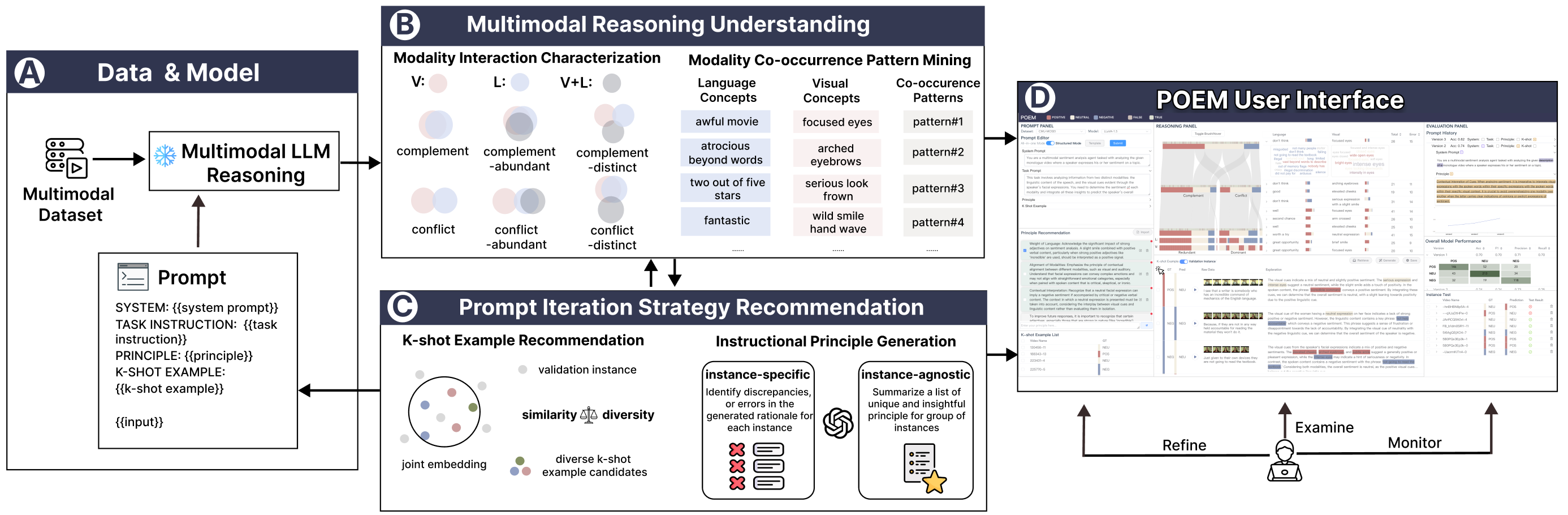}
    \caption{The \name system framework comprises four primary modules. (A) The visual and language modality information from the multimodal video dataset is processed by expert models, which are then fused and fed into multimodal LLMs. (B) The multimodal reasoning understanding module summarized the nuanced modality interactions and patterns at global and group levels. (C) The prompt iteration strategy recommendation panel provides diverse support for prompt refinement with semi-automatic k-shot example construction and instructional principle generation. (D) The \name interface facilitates efficient prompt performance examination, prompt refinement assistance, and prompt monitoring and comparison.
sys    }
    \label{fig:system_workflow}
\end{figure*}

Our goal is to develop a visual analytics approach that \jianben{streamlines} prompt engineering, empowering model practitioners to \jianben{efficiently} adapt and steer the multimodal reasoning performance of LLMs for targeted downstream tasks.
\jianben{By systematically} understanding how models integrate multimodal information for reasoning, users can evaluate and enhance knowledge in underperforming areas through proper prompt design \jianben{informed by domain expertise.}
To better understand users' requirements for system design, we worked closely with four experts in NLP and multimodal machine learning (\textbf{E1-E4}, \textbf{E1} is the coauthor).
\textbf{E1} is a researcher specializing in developing interactive systems for NLP and multimodal model analysis.
\textbf{E2} is an industry researcher responsible for applying and developing multimodal models for real-world applications.
\textbf{E3} and \textbf{E4} are Ph.D. candidates with multiple top conference publications in the areas of multimodal machine learning and multimodal LLMs. 


All experts concurred that there is a lack of tools for systematically analyzing the multimodal reasoning performance of LLMs. 
The current practice typically begins with observing the model's overall performance metrics, followed by randomly sampling instances to examine reasoning correctness. 
While few datasets~\cite{lu2022scienceqa, lian2023explainable} provide expert-written rationales as ground truth, the intricate interplay across modalities and extensive variability in free-text expressions makes it challenging to systematically understand the knowledge models use for reasoning and pinpoint their weaknesses.
Moreover, crafting and refining prompts to effectively elicit the desired knowledge \jianben{from} models for specific tasks often require labor-intensive and tedious prompt iterations. 
Consequently, an integrated tool is desired to facilitate systematic investigation of model behaviors at various levels and support well-informed prompt iterations with less cognitive effort.
The design requirements are summarized as follows:
\begin{enumerate}[label=\textbf{R{\arabic*}}, nolistsep]
\item \textbf{Summarize the impact of prompts on multimodal reasoning performance across varying levels of detail}
\jianben{When evaluating the reasoning performance of different prompts, users focus not only on overall statistics but also on how well the model's reasoning aligns with established knowledge at group and instance level.}
Therefore, it is crucial for the system to support multi-level and multi-faceted investigation of the model's multimodal reasoning performance.
Initially, the system should present a global overview of model performance. \jianben{As \textbf{E1} noted, ``understanding how different modalities interact is crucial for interpreting the model's behavior in the context of multimodal reasoning.''}
Users need to recognize the modalities the model relies on for its decisions and how the model behaves when different modalities present complementary or contradictory information.
After \jianben{gaining} a global understanding, users also need \jianben{insights into} how evidence from distinct modalities and their combinations influence the model. For example, \textbf{E3} expressed interest in identifying which types of visual cues or spoken words the model interprets as key indicators during reasoning. Besides, users need to inspect the model's output at the instance level to intuitively understand and \jianben{verify} the alignment of rationales with the original data.

\item \textbf{Provide comparative analysis of different prompt performance}
Multiple aspects of prompts influence model reasoning performance, including the structure and content of \jianben{task-specific instructions}, as well as the choice and order of demonstration examples.
\jianben{Navigating and exploring the evolving dynamics of prompts is necessary for users to ``identify influential factors for improvement'', as \textbf{E2} commented.
Therefore, it is imperative for the system to document prompt alternations, support streamlined prompt testing, and assist users in tracking and comparing the effects of diverse prompt modifications throughout the refinement process.}
This process facilitates an understanding of how different modifications impact model reasoning performance, thereby offering valuable insights for users to provide appropriate feedback and make informed decisions regarding subsequent iterations.


\item \textbf{Facilitate effective prompt refinement in diverse and efficient manner}
After pinpointing areas of underperformance, users can align and elicit model knowledge through refining prompts. 
This refinement \jianben{includes} providing more precise \jianben{task and scenario descriptions}, clear outlining principles for the model to follow, and supplementing with informative \jianben{demonstration} examples that help the model grasp the intended relationships.
However, given the vast range of potential feedback options, 
it imposes a huge cognitive burden on users to manually revise task articulation, formulate principles from scratch, and source the most informative examples for learning.
Furthermore, since the feedback users intend to provide often stems from their intuition and expertise, encompassing both inductive and deductive reasoning~\cite{petridis2023constitutionmaker}, the system ought to assist in translating these intuitive insights into concrete prompt content in an efficient and user-friendly manner.
For instance, \textbf{E4} suggested providing diverse prompt templates for easy selection. \textbf{E1} emphasized the need for a feature 
 that converts users' fragmented feedback into systematic principles for the model to follow, while \textbf{E3} highlighted the importance of automatically sourcing informative examples to provide high-quality rationales for knowledge alignment.






\end{enumerate}

\section{System \& Methods}
\label{sec: system and methods}
We designed \name based on the distilled design requirements in ~\cref{sec: design requirements}. In this section, we first introduce the overall system framework. Then we illustrate the methods for data processing, multimodal rationale understanding, and prompt iteration strategy recommendation.

\subsection{System Framework}
\Cref{fig:system_workflow} demonstrates the overarching workflow of the system. The multimodal video dataset, \jianben{processed into image frames (visual modality) with spoken narratives (language modality), along with the prompt, serves as input for the multimodal LLM.}
The multimodal LLM then performs reasoning and generates free-text answers for each input instance. (\cref{fig:system_workflow}A). Subsequently, the Multimodal Reasoning Understanding module (\cref{fig:system_workflow}B) provides a multi-level analysis of the generated free-text answers for a systematic understanding of the model's reasoning behavior.
Initially, it characterizes different interaction types between modalities. 
Then, a multimodal reasoning pattern mining algorithm is employed to identify intricate and fine-grained reasoning patterns. 
Concurrently, the Prompt Iteration Strategy Recommendation module (\cref{fig:system_workflow}C) offers varied support, including bottom-up k-shot example recommendations that balance similarity and diversity, and top-down instructional principle summarization at both instance-specific and agnostics levels aided by an auxiliary LLM. 
This module is designed to facilitate efficient prompt refinement, aiming to elicit and enhance specific knowledge to guide and improve model performance 

In the \name interface (\cref{fig:system_workflow}D), users have the option to either input their own prompts or choose from available templates in the \textit{Prompt Panel}. Subsequently, they can inspect the model's multimodal reasoning performance from different levels of detail. Specifically, at a global level, users can inspect the model's overall performance in the \textit{Evaluation Panel} and the interaction between and within modalities in the \textit{Reasoning Panel}. At the group level, users are able to scrutinize the model's reasoning patterns concerning different concepts spanning across modalities. At the instance level, users can examine individual instances in detail for verification.
Users can then revise and incorporate principles and/or k-shot examples into prompts based on automatic recommendations and insights obtained from the current model and prompt performance examinations. The refined prompt can then be sent to the model for evaluation in the \textit{Prompt Panel}.
In the \textit{Evaluation Panel}, users can evaluate and compare the effect of each prompt iteration on both global model performance and individual instances. Additionally, they can monitor and track detailed changes across various prompt versions and iteratively refine the prompt to achieve satisfactory multimodal reasoning performance.

\subsection{Dataset and Model}
\label{sec: data and model}
We demonstrate the effectiveness of our system on two different datasets for multimodal content comprehension tasks: CMU-MOSEI~\cite{zadeh2018multimodal} for multimodal sentiment analysis, and WTaG~\cite{bao2023foundation} for user intent understanding.
The CMU-MOSEI~\cite{zadeh2018multimodal} dataset consists of monologue video clips in which speakers express their sentiments about a specific topic. 
The WTaG dataset~\cite{bao2023foundation} comprises egocentric video clips of users performing cooking tasks under the guidance of an instructor within an augmented reality setting.
The videos within both datasets contain information from two primary modalities: the language modality, represented by spoken content, and the visual modality, characterized by the scenes and user behaviors depicted in the videos. 
\jianben{Both datasets include ground-truth labels for evaluation.}
Following the practice in prior works~\cite{mishra2023promptaid, sherry2023scattershot}, we split each dataset into three subsets: a validation set, a demonstration example set, and a test set.
In the splitting process, we ensure that the label distribution remains consistent across these subsets. 
The validation set serves the purpose of prompt iteration evaluation. The demonstration example set facilitates the construction of k-shot examples, and the test set provides additional instances beyond the validation set for a more comprehensive assessment of prompt efficacy. 
The size of the validation set needs to be moderate so that users can get timely feedback during the prompt iteration while also covering diverse data patterns for comprehensive model reasoning performance diagnosis. Based on our preliminary experiment, we maintain a distribution ratio of 1:2:1 for the validation, demonstration, and test sets, respectively.
We also implemented batch processing to improve the system's response speed.


\jianben{Regarding the model setting, we employ the LLaVA~\cite{liu2023visual} and GPT-4V(ision)~\footnote{https://openai.com/index/gpt-4v-system-card/} model to perform multimodal reasoning considering their strong reasoning and instruction-following abilities. We specifically utilize the “llava-v1.5-13b” and “gpt-4-vision-preview” version.
It's important to note that our approach is designed to be model-agnostic, meaning other multimodal LLMs that support multimodal content reasoning, such as Gemini~\footnote{https://deepmind.google/technologies/gemini/} and LLaMA series~\cite{touvron2023llama}, can be easily integrated into the system.}
\jianben{For each video clip, we followed the commonly adopted practice~\cite{fu2024video, yang2023mmbigbench}, sampling frames per second to compose an image sequence from the visual modality, which is then combined with the corresponding spoken content from the language modality as input of the multimodal LLM.}

\jianben{The input prompt for the multimodal LLM reasoning follows the general prompt structure $(I,\left\{x_i, y_i\right\}_{i=1}^k, x_t)$~\cite{kim2023evallm, wei2022cot, wu2022promptchainer}. }
Here, $I$ is the task-specific instructions elaborating on the targeted scenarios, tasks to be finished, and expected output structure (e.g., return your answer in a JSON object).
To propel the LLM to perform CoT reasoning, The prompt could include instructions like ``Please provide a step-by-step analysis''.
$\left\{x_i, y_i\right\}_{i=1}^k$ is the demonstration example set. Each demonstration example includes input $x_i$ and output $y_i$, \jianben{where $x_i$ follows the same format as the validation set and $y_i$ includes the correct rational and final answer as provided by the ground-truth labels.}
We also support the zero-shot setting where demonstration examples are not provided. 
Finally, for each test input $x_t$, the LLM is expected to generate the output $y_t$, containing a free-text \jianben{rationale} and a \jianben{final answer}.

\subsection{Multimodal Rationale Understanding}
\label{sec: multimodal rationale understanding}
\subsubsection{Modality Interaction Characterization}
\jianben{Understanding how multimodal models utilize information from distinct modalities and integrate it to make cross-modal inferences is crucial for gaining insight into the model's reasoning performance.}
Several works~\cite{wang2022m2lens, liang2023multiviz} have tried to characterize the interaction between different modalities based on aggregated feature attribution values~\cite{ribeiro2016lime, lundberg2017shap}. There are also works~\cite{liang2023quantifying, liang2023multifusion, liang2024quantifying, yu2023mixture} trying to quantify the degree of interactions between modalities with a partial information decomposition framework. 
Building on the foundation of these works, we characterize the modality interaction in the context of our targeted multimodal \jianben{reasoning} tasks as follows:

Considering the labeled multimodal dataset with two modality $\mathcal{X}_1$ and  $\mathcal{X}_2$, the unimodal data $\mathcal{D}_i=\left\{\left(x_i, y\right):\mathcal{X}_i \times \mathcal{Y}\right\}$ where $i \in\{1,2\}$
, and the multimodal data $\mathcal{D}_M= \left\{\left(x_1, x_2, y\right): \mathcal{X}_1 \times \mathcal{X}_2 \times \mathcal{Y}\right\}$. 
\jianben{When performing chain-of-thought reasoning, for each input data point, the output of the multimodal LLM includes a free-text rationale and a final answer.}
\jianben{Here, we denote the sample space where the multimodal LLM performs reasoning using information from a single modality as $ f_i: \mathcal{X}_i \rightarrow \Delta \mathcal{Y}_i$ and the sample space where the multimodal LLM performs reasoning with information from both modalities as $ f_M: \mathcal{X}_1 \times \mathcal{X}_2 \rightarrow \Delta \mathcal{Y}_M$. where $\Delta \mathcal{Y}$ denotes the probability simplex of final output answer.}
\jianben{For $f_a$ and $f_b$ where $a, b\in\{1,2, M\}$,
The distance function can be defined as $d\left(f_a, f_b\right)= \left\|\Delta_a-\Delta_b\right\|$ to measure the distance between $f_a$ and $f_b$.}
Based on the distance function, we can define two basic interaction types between two modalities.
When $d\left(f_1, f_2\right)< \theta$, where $\theta$ is a pre-defined threshold, the interaction type is \textbf{complement}, indicating these two modalities 
contribute to the final answer in the same direction.
Conversely, when $d\left(f_1, f_2\right)> \theta$, the interaction type is \textbf{conflict}, indicating these two modalities provide discrepant information for reasoning. 
By further considering how the final answer will change when analyzing information from each modality independently, and combining information from two modalities jointly for reasoning, i.e., the distance function $d\left(f_1, f_M\right)$ and  $d\left(f_2, f_M\right)$, we can define subdivided interaction type~\cite{liang2023quantifying, liang2024quantifying} as shown in ~\autoref{fig:system_workflow}B:
\begin{itemize}[itemsep=0pt,topsep=0pt,parsep=0pt]
\item ~\textbf{Complement-Redundant}: when $f_M$ aligns with $f_1$ and $f_2$, where $d\left(f_1, f_M\right) < \theta$, and $d\left(f_2, f_M\right) < \theta$ 

\item ~\textbf{Complement-Distinct}: when $f_M$ distinct from $f_1$ and $f_2$, where $d\left(f_1, f_M\right) > \theta$, and $d\left(f_2, f_M\right) > \theta$ 

\item ~\textbf{Conflict-Dominant}: when $f_M$ aligns with $f_1$ or $f_2$, where $d\left(f_1, f_M\right) < \theta$ and $d\left(f_2, f_M\right) > \theta$ or switch the $f_1$ and $f_2$

\item ~\textbf{Conflict-Distinct}:  when $f_M$ distinct from $f_1$ or $f_2$, where $d\left(f_1, f_M\right) < \theta$, and $d\left(f_2, f_M\right) < \theta$ 
\end{itemize}
In this paper, we primarily focus on the visual and language modalities, which are the main subjects of investigation in current multimodal LLM research. For more modalities, the interaction characterization framework can be extended by pairwise comparison.

\subsubsection{Multimodal Reasoning Pattern Mining}
\label{sec: multimodal reasoning pattern mining}
Upon gaining insight into the model's reasoning process at the modality interaction level, it becomes crucial to identify the specific concepts or their combinations within and across individual modalities the model utilizes for reasoning.
\jianben{As shown in \cref{fig:system_workflow}B},
we parse the generated \jianben{rationale} into a list of intermediate evidence along with their associated inferences that contribute to the final answer. For example, within a free-text \jianben{rationale} generated by the LLM, \jianben{\textit{``The serious expression suggests a neutral sentiment, while in the spoken content, the phrase 'incredible command' conveys a positive sentiment.} The visual evidence ``serious expression'' infers ``neutral'' sentiment and the language evidence ``incredible command'' implies ``positive'' sentiment.}
Given the LLM's generative characteristics resulting in the variability of evidence across different \jianben{rationales}, we employed the \textit{text-embedding-3-small}\footnote{https://platform.openai.com/docs/guides/embeddings} model to calculate embeddings for all extracted evidence (e.g., ''serious expression'' and ''incredible command'' et al). Subsequently, we utilized the HDBSCAN algorithm~\cite{mcinnes2017hdbscan} to cluster visual and language evidence respectively. We identified the evidence located closest to the cluster's centroid as the representative concept for each cluster. Subsequently, we utilized the Apriori~\cite{agrawal1994fast} algorithm to identify frequent patterns of concept co-occurrence within and across different modalities in the generated rationales \jianben{for validation set}.
This approach enables users to conduct a more structured and \jianben{comprehensive} analysis of the patterns within the generated rationales, allowing them to identify potential recurring biases or errors made by the model.

\subsection{Prompt Iteration Strategy Recommendation}
\label{sec: prompt iteration strategy recommendation}
As mentioned in ~\cref{sec:related_works}, the content and phrasing of task instructions, along with the choice of demonstration examples, can greatly influence the model's reasoning performance. 
Our preliminary experiment and expert interview results suggested that the instruction content (e.g., task specifications) and the choice of demonstration examples exert a more pronounced effect on model performance than the precise wording used for our targeted multimodal reasoning tasks.
Therefore, in this paper, we mainly focus on facilitating users in instruction content refinement and demonstration example construction.

\subsubsection{K-shot Example Recommendation}
Few-shot prompting has been a data-efficient strategy to adapt LLMs for specific downstream tasks using merely a handful of illustrative input-output pairs.
However, the effectiveness heavily relies on the choice of examples to inform the model about the desired mapping~\cite{sherry2023scattershot, jiang2022promptmaker}.
Identifying informative examples for effectively guiding the model can be challenging for users. 
Moreover, beyond simply pairing inputs with final answers, reasoning necessitates providing a rationale for each example, which is equally difficult for users to craft on their own. 


To enhance the efficiency of sourcing the demonstration example set, we first employed the k-nearest neighbors algorithm to sample the candidate k-shot example set considering both relevancy and diversity. 
\jianben{For each instance in both the validation set and demonstration example sets, we computed embeddings for the visual (images) and language (text transcript) modalities separately and then concatenated these embeddings to represent each instance. 
Specifically, we utilized the pre-trained CLIP\footnote{https://huggingface.co/sentence-transformers/clip-ViT-B-16} model, which maps text and images to a shared vector space for embedding computation. Further details are provided in the supplementary material.}
\jianben{For each validation instance, we identified its k-nearest neighbors as potential candidates based on their embedding cosine similarity.
These candidates were then ranked in descending order of similarity.}
To select the final k-shot examples, we prioritized both ranking and label diversity, ensuring the inclusion of all possible labels in the final set to prevent model bias. 
To streamline the process of crafting rationales for users, we integrated the \textit{gpt-4-turbo model}~\footnote{https://platform.openai.com/docs/models/gpt-4-turbo-and-gpt-4} to automatically generate structured rationales for each demonstration example based on its ground truth labels. 
This approach offers users a preliminary basis for refinement, sparing them the need to begin from scratch. 
Furthermore, we utilized the refinements operated by users to iteratively enhance the quality of the generated rationales. 
These demonstration examples are then combined into the sequence$\left\{x_i, y_i\right\}_{i=1}^k$ for inclusion in the prompt, where $y_i$ is the rationale and final answer provided by users for $x_i$~(\autoref{fig:system_workflow}C).




\subsubsection{Instructional Principle Generation}
While k-shot examples aim to inductively teach the model the correct mappings between input-output pairs, providing explicit principles regarding proper practices or clarifying potential errors has also proven to be an effective strategy for drawing out desired knowledge and guiding model performance~\cite{zhang2024context, petridis2023constitutionmaker}. 
Humans generally formulate principles in two ways. One involves directly leveraging their existing knowledge. For example, the principle for identifying sarcasm could be to ``pay attention to the inconsistency between a word's literal interpretation and its contextual meaning.'' 
The other is that individuals derive lessons from specific instances and subsequently aggregate these instance-level insights into higher-level principles in a bottom-up manner.
However, users may find it difficult to immediately generate principles from scratch, derive insights by manually examining instances one at a time, and fully articulate their principles considering the complexity of multimodal reasoning.
\jianben{For this purpose, we employed an auxiliary LLM to facilitate the summarization and recommendation of principles. We selected the \textit{gpt-4-turbo model} for its strong capabilities in text understanding and summarization, and it can be replaced by more advanced models in the future.}

Specifically, we instructed the \textit{gpt-4-turbo model} to produce principles at both instance-specific and instance-agnostic levels (\autoref{fig:system_workflow}C). 
At the instance-specific level, the model is tasked with analyzing discrepancies between generated reasoning and ground truth answers for each instance, summarizing potential error causes, and further deriving principles to avoid similar mistakes. 
At the instance-agnostic level, We instruct the model to condense the generated instance-specific principles into more generic principles tailored to the specific targeted task.
It is important to acknowledge that the generated principles may not always be accurate and should not be treated as golden rules. Their primary purpose is to provoke thought and inspire users to conceive new ideas or enhance existing ones rather than initiate from zero.
Thus, users are empowered to either input and create their principles or choose to amend and revise principles that have already been generated according to their preferences. 
\jianben{Details regarding the prompt used for principle generation are provided in the supplementary material.}






\section{Interface Design} 
The \name interface (\cref{fig:teaser}) consists of three coordinated views to assist users in seamlessly evaluating the impact of different prompts, refining prompts through semi-automatic suggestions, and conducting iterative testing of prompts. In this section, we introduce the design of each view and the interactions that connect them in detail.

\subsection{Prompt Panel}
The \textit{Prompt Panel} (\cref{fig:teaser}A) provides flexible prompt operations to support smooth prompt engineering experience (\textbf{R3}). 
Upon selecting the dataset and model, users can craft the prompt on their own or initiate by selecting from a list of prompt templates collected from state-of-the-art benchmarks~\cite{yang2023mmbigbench, fu2024video} in the \textit{Prompt Editor} (\cref{fig:teaser}A-1). 
The prompt is organized into distinct sections, as introduced in ~\cref{sec: data and model}, to facilitate a clear and straightforward editing experience. Users can also switch to the plain text editing mode for editing and format checking before submission.
The \textit{Principle Recommendation} view (\cref{fig:teaser}A-2) displays an organized summary of principles for user validation. 
The generated instance-specific and agnostic principles are differentiated by background colors: gray for instance-specific principles and green for instance-agnostic principles.
Newly generated principles are marked with red dots at the top right for highlighting.
Users can modify any existing principle by utilizing the editing function or articulate their principles via the principle input box. Furthermore, users are allowed to delete any principles deemed inappropriate or redundant. Subsequently, upon selecting the desired principles, users can integrate them into the current prompt within the prompt editor by clicking the ``Import \includegraphics[width=0.015\textwidth]{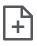} '' button. 
The \textit{K-shot Example List} (\cref{fig:teaser}A-3) below provides a concise summary of K-shot examples with user-annotated rationales, waiting for further editing or inclusion into the prompt.

\subsection{Reasoning Panel}
The \textit{Reasoning Panel} (\cref{fig:teaser}B) facilitates a thorough investigation of the model's multimodal reasoning behaviors, from global and sub-group patterns down to specific individual instances (\textbf{R1}).

A three-layer Sankey diagram-based design (\cref{fig:teaser}B-1) is adopted to portray interactions among modalities at the global level. 
The first layer demonstrates the overall distribution of prediction classes and errors with two vertically stacked barcode charts. The horizontal length encodes the number of instances, and the color encodes the corresponding class and error. Instances belonging to the same class are positioned close together to enable easy exploration both within and between classes.
The second intermediate layer summarizes the conflict and complement relationship between visual and language modalities and adopts the same encoding as the first layer. 
The third layer delves into the fine-grained four types of modality interactions. While retaining the same visual encoding for the prediction class and error distribution in each interaction type, this layer introduces two additional barcode charts to delineate the prediction result of the single visual and language modality, thus illustrating the detailed distributions of the two modalities across various types of interactions. 
Besides, two adjacent layers are interconnected through flows, the width of which is proportional to the number of instances they encompass. Hovering over the flows will highlight the related instances across all three layers.
Users can also brush the barcode chart in each layer to select an interested group of instances for further investigation. \jianben{The selected instances will be highlighted with a grey background.} \jianben{The corresponding mined patterns and instances will be displayed on the right and below respectively.}

After selecting the interested group of instances, the multimodal reasoning pattern mining algorithm in ~\cref{sec: multimodal rationale understanding} is applied, with the extracted patterns displayed in the table on the right (\cref{fig:teaser}B-2). Each row exhibits one distinct pattern with its representative visual and language concepts, support (i.e., contained instance numbers), and error statistics. They can sort and filter the patterns based on these statistics by clicking on the corresponding column. The representative language and visual concepts are shown for intuitive pattern understanding by users. Adjacent to each concept, a stacked bar chart presents the distribution of its associated class. 
Users can expand each row to view the detailed distribution of evidence in a word cloud, where each phrase's size represents its frequency of occurrence and its color denotes the proportion of associated classes.
Users can select patterns or evidence of interest by clicking, and the corresponding instances will be displayed in the instance view below.

The instance view below (\cref{fig:teaser}B-3) is designed to expedite the examination and verification of individual instances by showcasing the original multimodal video content along with its detailed reasoning. The raw data column exhibits the video's keyframe image sequence and spoken content to enable quick visual and language content digestion and \jianben{validation}. Users can hover over these frames for an enlarged view and playback the original video for rapid verification.
Subsequent columns present the ground truth labels, the model's predictions, and the generated free-text rationales. To enhance readability and quick text comprehension, evidence is highlighted with the corresponding color of its associated class. The ground truth and prediction columns are also colored for easy comparison.
Users can select instances for principle generation by clicking the ``Generate \includegraphics[width=0.015\textwidth]{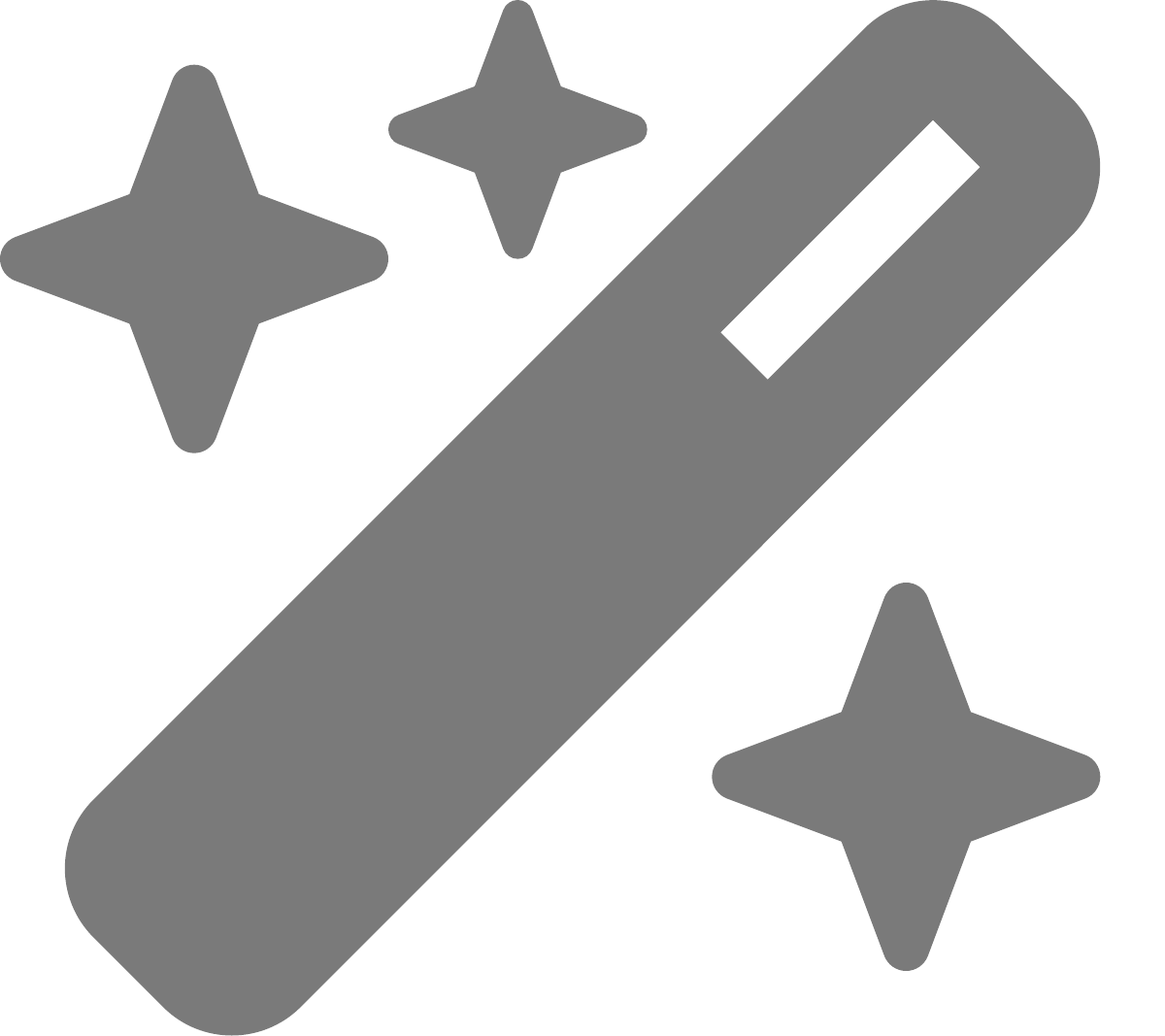} '' button, after which the generated principles will be listed in the \textit{Principle Recommendation} view.
In addition to displaying selected validation instances for review, users can toggle to the \textit{K-shot Example Mode}(~\autoref{fig: casetwo}A). Within this mode, the interface presents the ranked list of k-shot examples recommended by the proposed sampling strategy in ~\cref{sec: prompt iteration strategy recommendation}. For each example, the interface details its raw data (i.e., keyframe sequence and spoken narratives in the raw video), its ground truth, and the rationales. Users can modify the content in the corresponding column directly to provide high-quality rationales. Moreover, users can source more k-shot examples by clicking the ``Retrieve \includegraphics[width=0.015\textwidth]{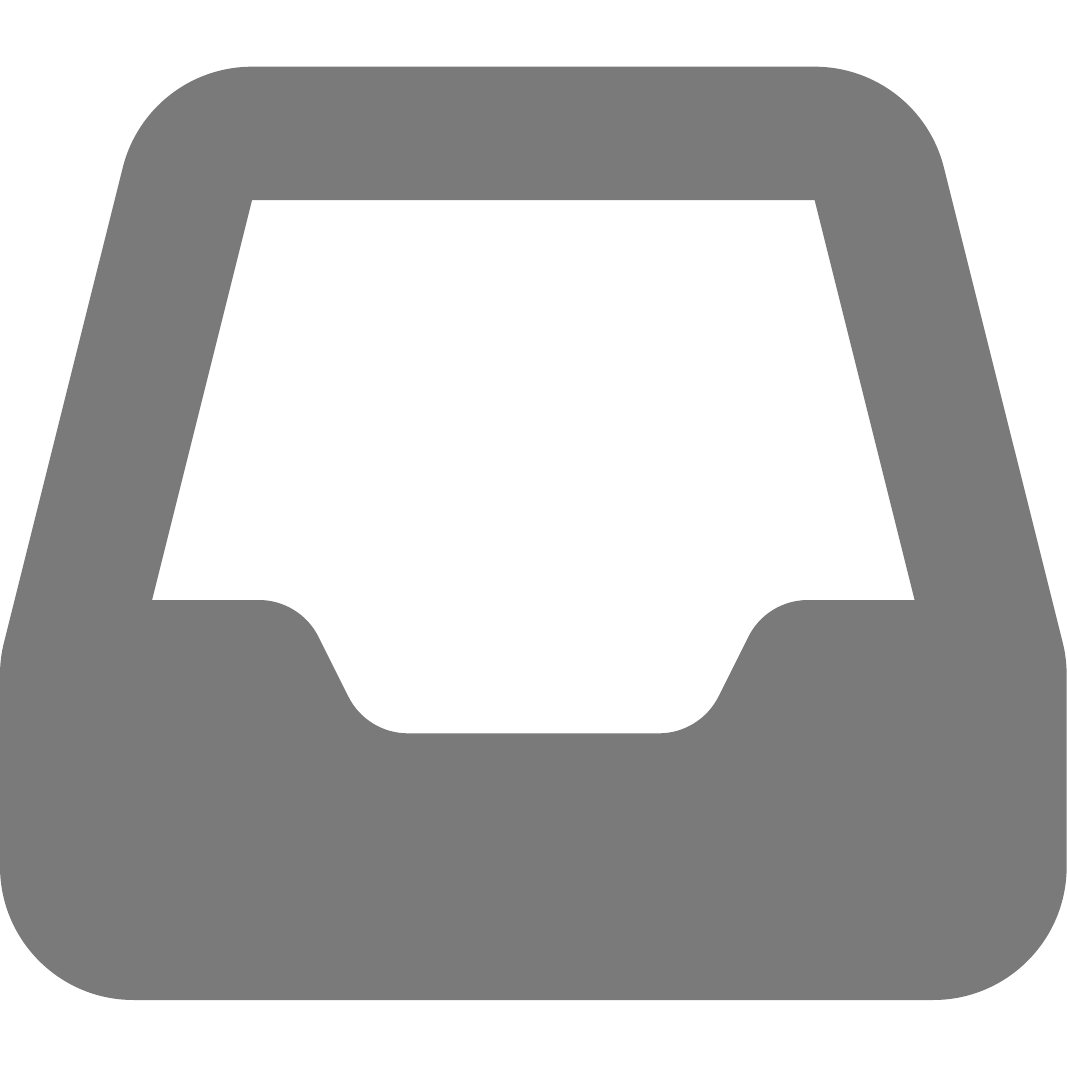} '' button and save the selected ones to the \textit{K-shot Example List} with the ``Save \includegraphics[width=0.015\textwidth]{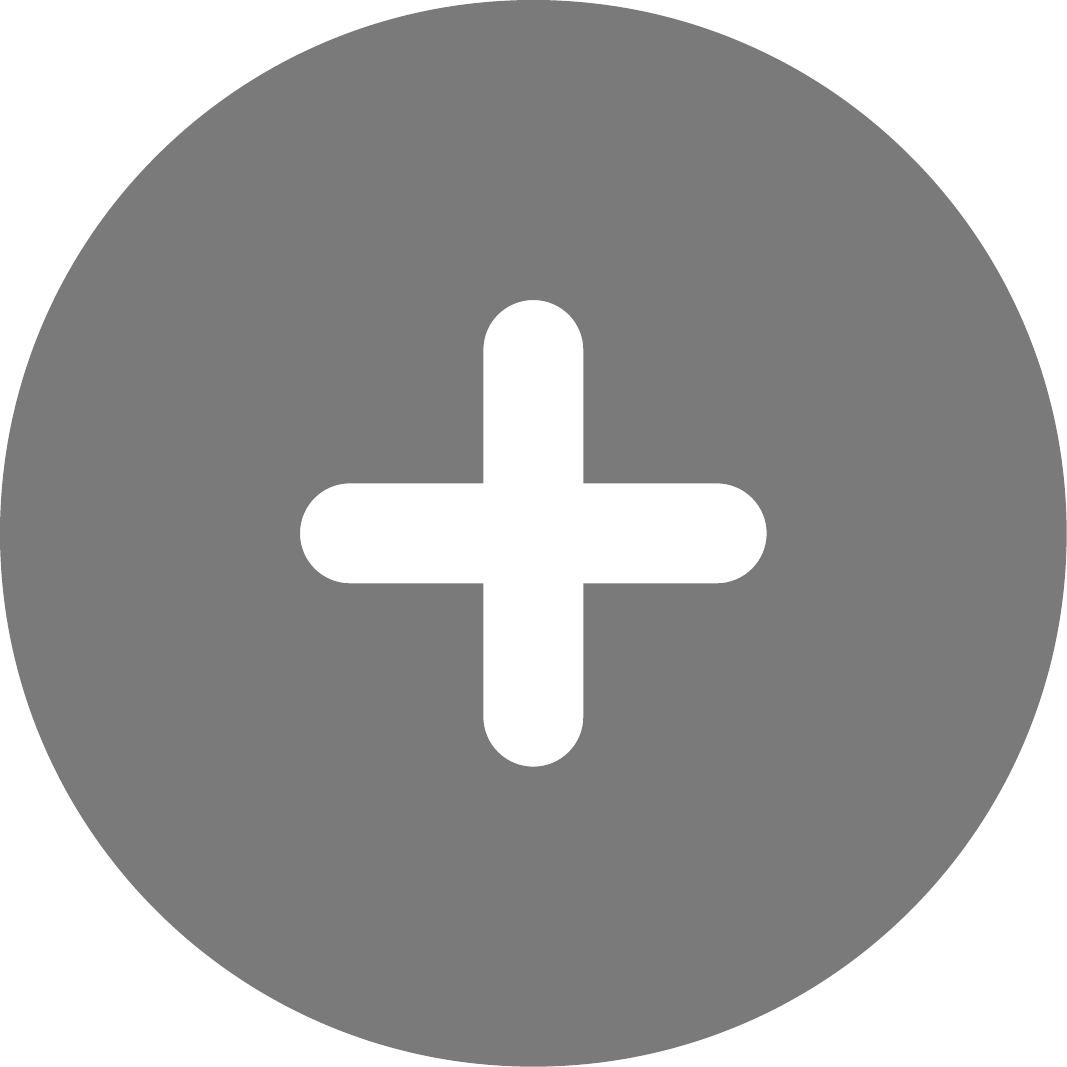} '' button.

\subsection{Evaluation Panel}
The \textit{Evaluation Panel} (\cref{fig:teaser}C) offers comprehensive insights into both global and local performance of prompts, along with the prompt iteration history for efficient monitoring and comparison of prompt performance (\textbf{R2}).

The \textit{Prompt History} view (\cref{fig:teaser}C-1) archives previous prompts regarding their content and performance. Each row represents a prompt version with its accuracy and modifications are summarized using intuitive icons. This design enables users to easily compare performance and trace alterations in different sections of the prompts.
Users can expand and collapse each row for a hierarchical examination of modifications within each prompt section. 
Detailed additions and deletions in the content are distinctly marked and highlighted through varied colors and line styles.
\jianben{The line chart below shows the model accuracy change.}
The \textit{Overall Model Performance} view (\cref{fig:teaser}C-2) records the global performance statistics of each prompt iteration. 
Users can expand each row to inspect the detailed confusion matrix.
The \textit{Instance Test} view (\cref{fig:teaser}C-3) exhibits the performance of prompts on individual instances that are of particular interest to users.
Users can select instances from \textit{Reasoning Panel} and save them to observe their performance change during prompt iterations with the ``Save \includegraphics[width=0.015\textwidth]{figs/save-button.pdf} '' button. 
They can also source additional unseen test instances with the ``Retrieve \includegraphics[width=0.015\textwidth]{figs/retrieve-button.pdf} '' function.

\section{Evaluation}
In this section, we showcase the efficacy and efficiency of \name via two case studies and feedback gathered from expert interviews. 
The primary objective of the two case studies is to help users obtain well-performing prompts utilizing their domain expertise and knowledge to guide LLM's multimodal reasoning performance with minimal effort.

\subsection{Case One: Improving multimodal sentiment reasoning with CMU-MOSEI dataset}
\textbf{E5}, a sentiment analysis expert, seeks to generate effective prompts for steering LLM's multimodal sentiment reasoning performance with the CMU-MOSEI dataset. The LLM is tasked with interpreting the speakers' verbal and visual signals to determine their sentiment as ``positive'', ``negative'', or ``neutral''.

\begin{figure}[ht]
    \centering
    \includegraphics[width=1.0\linewidth]{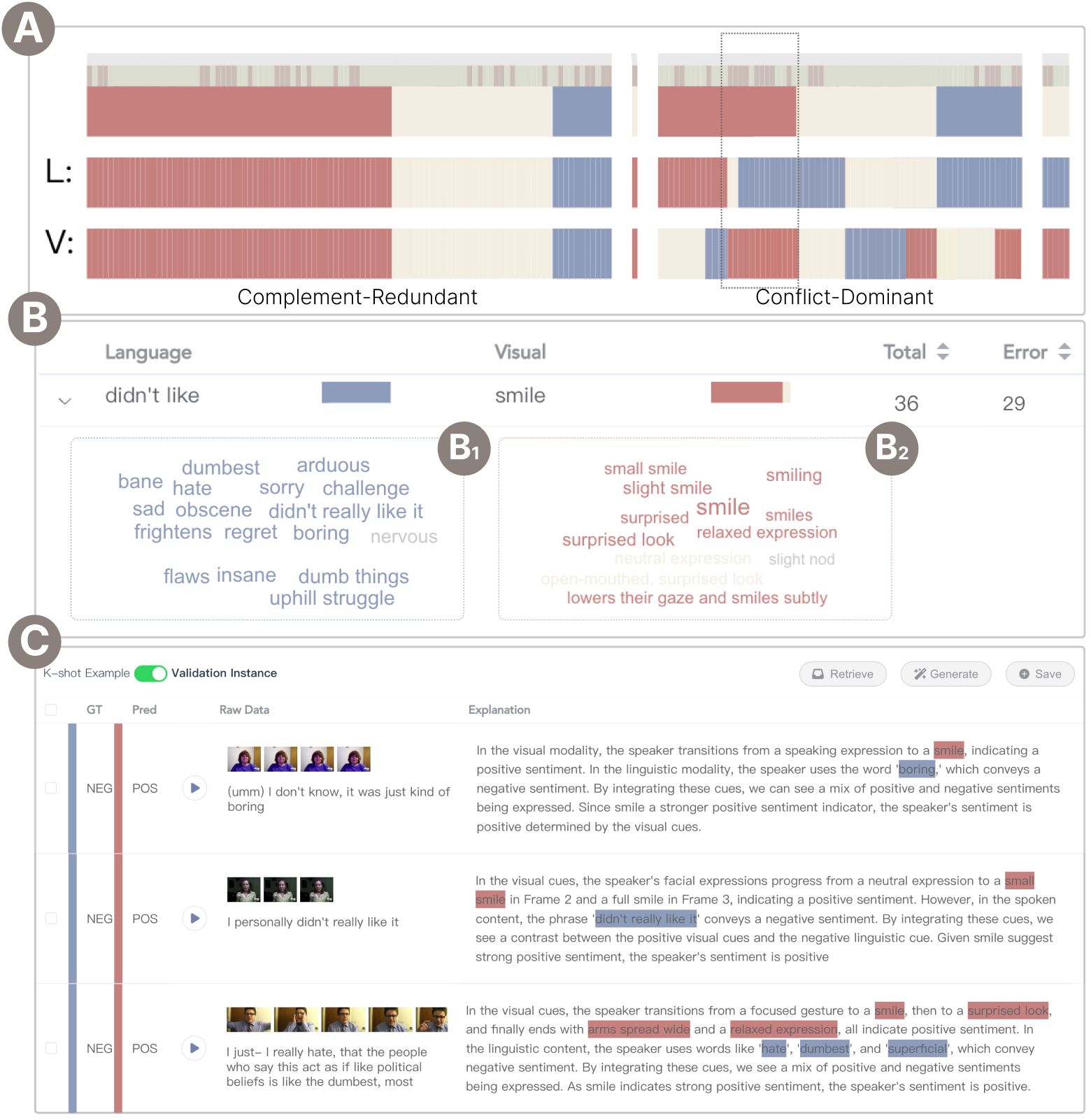}
    \caption{(A) Identified dense error areas in \textit{conflict-dominant} modality interaction. (B) The multimodal pattern ``didn't like'' and ``smile'' and their associated evidence group. (C) The error cases where ``smile'' predominated and biased the reasoning process.}
    \label{fig: caseone_first}
\end{figure}

After loading the dataset and model, \textbf{E5} initially selected and submitted a provided prompt template in the \textit{Prompt Panel} to evaluate its performance (\textbf{R2}), \jianben{which yielded an accuracy of 70\%.}
To gain an overview of the interactions between the visual and language modalities in the LLM's reasoning process (\textbf{R1}), \textbf{E5} began by examining the Sankey diagram in the \textit{Reasoning Panel}.
\jianben{Through observing the length and error distribution of the barcode charts in the first and second layers, \textbf{E5} noticed that the model tended to interpret sentiments as ``neutral'' or ``positive'' rather than ``negative''.}
Furthermore, in a large proportion of instances, the visual and language modalities provided complementary information, while in others, they presented conflicting information with increased errors.
\textbf{E5} was particularly interested in how the LLM reasoned in scenarios where the two modalities \jianben{presented} conflicting information and how errors occurred. So, she \jianben{explored} the third layer for a more fine-grained examination. 
At the third layer, she \jianben{identified} a dense \jianben{cluster} of errors within the \textit{conflict-dominant} relationship (\cref{fig: caseone_first}A), where the visual modality implied a positive influence, while the language modality suggested a negative one. The ultimate combined effect was positive, indicating that the visual modality \jianben{dominated} the reasoning process. 

Following this, \textbf{E5} brushed this group of instances to further inspect their contained reasoning patterns in the table on the right.
When going through the patterns \jianben{sorted} in descending order of error rate, \textbf{E5} discovered the combination of the language concept ``didn't like'' with the visual concept ``smile'' yielded high error rates (\cref{fig: caseone_first}B). 
The adjacent bar charts, predominantly \jianben{colored} in blue for ``didn't like'' and red for ``smile'', indicated that the LLM consistently interpreted language evidence concluded with ``didn't like'' as a negative signal and visual evidence \jianben{featuring} ``smile'' as positive during the reasoning process. 
She further explored this pattern \jianben{by unfolding the row}, where evidence under the language concept ``didn't like'' included phrases like ``arduous'', ``boring'', and ``hate'' highlighted in blue (\cref{fig: caseone_first}B-1), while the visual concept ``smile'' comprised instances such as ``small smile'', ``slight smile'', and ``smiling'' marked in red (\cref{fig: caseone_first}B-2).
\textbf{E5} thought these inferences for each modality reasonable but wondered how the correctly deduced evidence led to the final error. Therefore, she proceeded to inspect the detailed reasonings of individual instances exhibiting this pattern in the instance view below (\cref{fig: caseone_first}C).
Upon examining the raw data and the rationales generated by the LLM, \textbf{E5} figured out that the model correctly reasoned about individual modality, as in these instances, the speakers had explicitly stated their negative opinions verbally while \jianben{showing} mild positive facial expressions like gentle smiles. 
However, the LLM was biased by the positive visual cues, allowing them to overshadow and dominate its reasoning, despite the explicit negative sentiment conveyed through language.

\begin{figure}[ht]
    \centering
    \includegraphics[width=1\linewidth]{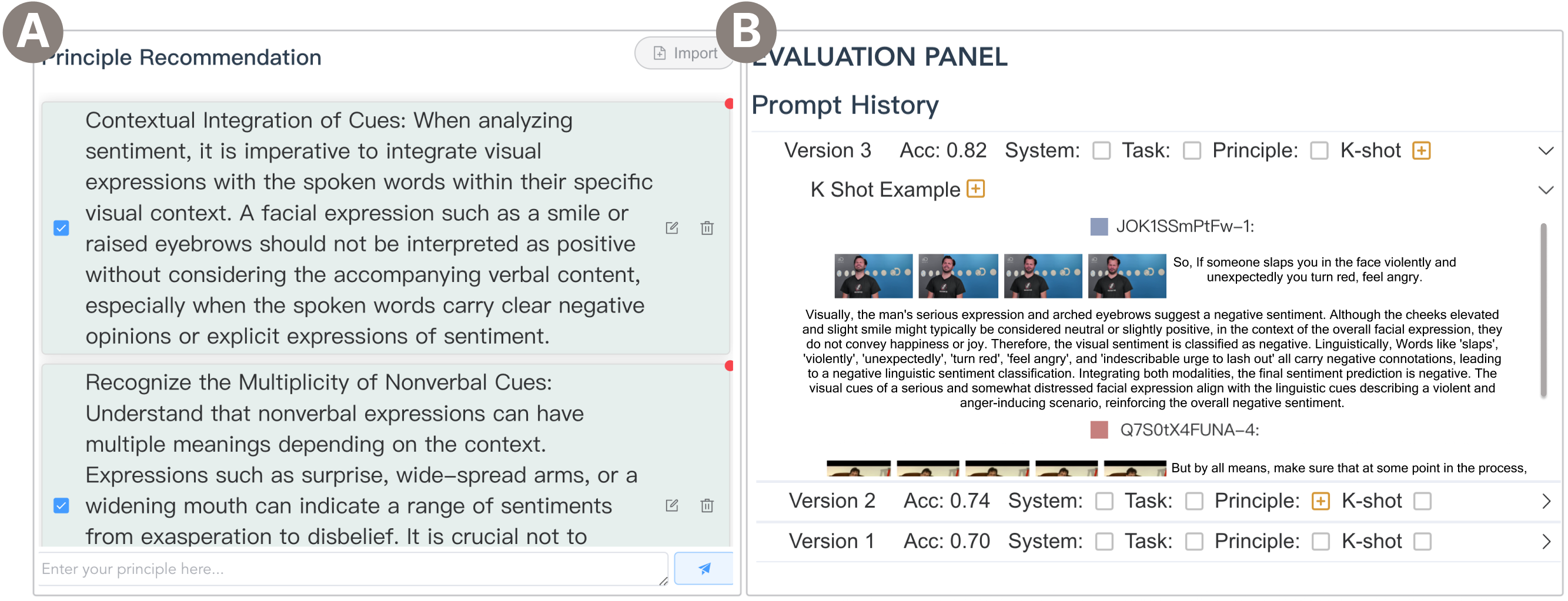}
    \caption{(A) The recommended principles for alleviating errors in case one. (B) The recorded prompt iteration history in case one.}
    \label{fig: caseone_second}
\end{figure}

Following this discovery, \textbf{E5} decided to derive principles from these erroneous cases to guide the LLM toward correct reasoning in this situation (\textbf{R3}). Therefore, \textbf{E5} selected these instances and clicked the ``Generate \includegraphics[width=0.015\textwidth]{figs/generate-button.pdf} '' button to generate principles. She also saved these instances of interest to the right \textit{test panel} for further validation. 
In the \textit{Principle Recommendation View}, \textbf{E5} reviewed the generated principles and identified well-articulated general principles that underscored the importance of interpreting visual cues alongside the corresponding verbal content with careful consideration of specific context (\cref{fig: caseone_second}A).
To ensure generalizability and avoid introducing new bias, \textbf{E5} \jianben{revised} the last sentence as ``\textit{It is crucial to avoid overemphasizing one modality over another when the latter carries clear indications of opinions or explicit expressions of sentiment.}''
Then, \textbf{E5} imported this principle into the prompt editor and submitted it for testing. In the \textit{Model Performance View}, she found a slight improvement in the overall accuracy from \jianben{70\%} to \jianben{74\%}. Meanwhile, in the \textit{Test Panel View}, she checked the performance of the new prompt on previously saved instances, the majority of which were now correctly reasoned (\autoref{fig:teaser}C).
This indicated that the incorporated principle had \jianben{effectively} guided the LLM to use the correct knowledge for reasoning in this scenario.

Subsequently, \textbf{E5} sought to enhance the model's reasoning stability and its ability to recognize varied patterns in sentiment analysis by incorporating some k-shot examples (\textbf{R3}). 
Thus, she switched to \textit{K-shot Example Mode}, where recommended K-shot examples with reasonings crafted by the auxiliary LLM were listed. \textbf{E5} selected the top three instances spanning distinct classes and refined the provided reasoning leveraging his knowledge and expertise. 
Upon completing the rationale annotations, \textbf{E5} appended these examples to the K-shot example list on the left side and imported them into the prompt. 
After running the test, the overall accuracy increased to 82\% (\cref{fig: caseone_second}B).

\subsection{Case Two: Enhancing Multimodal User Intention Understanding with WTaG dataset}
\textbf{E6} is an engineer tasked with building an intelligent virtual assistant to help users perform complex tasks within augmented reality environments.
Building such an assistant necessitates comprehending user intentions.
\textbf{E6} thus wanted to steer the GPT-4V(ision) model using \name to finish this task. 
\textbf{E6} experimented with the WTaG dataset~\cite{bao2023foundation}, \jianben{where the video clips were recorded from the user’s egocentric perspective.}
These clips included user-instructor dialogues captured by microphones and visual context encompassing the scene and user behaviors from head-mounted cameras.
The multimodal LLM needs to deduce the user's intention based on this multimodal context and categorize it into one of five classes: ``Question'', ``Answer'', ``Confirmation'', ``Hesitation'' and ``Self Description''. 

After initializing the dataset and model, \textbf{E6} first chose to use the prompt \jianben{provided} in the original dataset repository for validation (\textbf{R2}). The \textit{Evaluation Panel} revealed that this prompt \jianben{achieved only 53\%} accuracy in a zero-shot setting. While this result is higher than what was reported in the paper using \textit{gpt-3.5-turbo model}~\cite{bao2023foundation}, it is still insufficient for the task.
\textbf{E6} next examined the confusion matrix and noticed that the model's predictions were heavily biased towards the `` Confirmation'' and ``Answer'' classes (~\autoref{fig: caseone_first}A).
Upon randomly inspecting the model-generated rationales alongside the raw data incorrectly classified in the \textit{Reasoning Panel}, \textbf{E6} observed that while the model could adequately describe and analyze both visual and spoken content, it struggled to comprehend the meaning of designated prediction classes, especially ``Self Description.'' This resulted in scarce predictions for this class and a bias towards more familiar classes such as ``Confirmation'' and ``Answer''.

\begin{figure}[ht]
    \centering
    \includegraphics[width=1\linewidth]{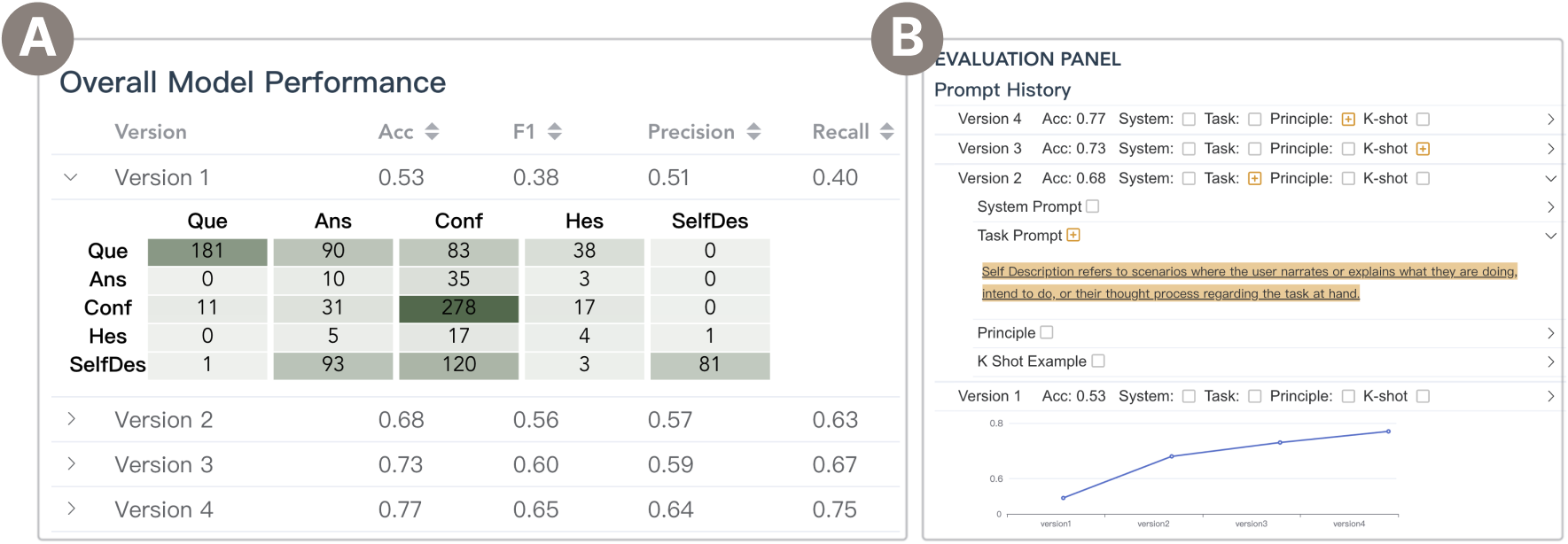}
    \caption{(A) The confusion matrix showing the model's prediction bias towards the ``Confirmation'' and ``Answer'' classes. (B) The recorded prompt iteration history in case two.}
    \label{fig: case_two_first}
\end{figure}

\begin{figure}[ht]
    \centering
    \includegraphics[width=1.0\linewidth]{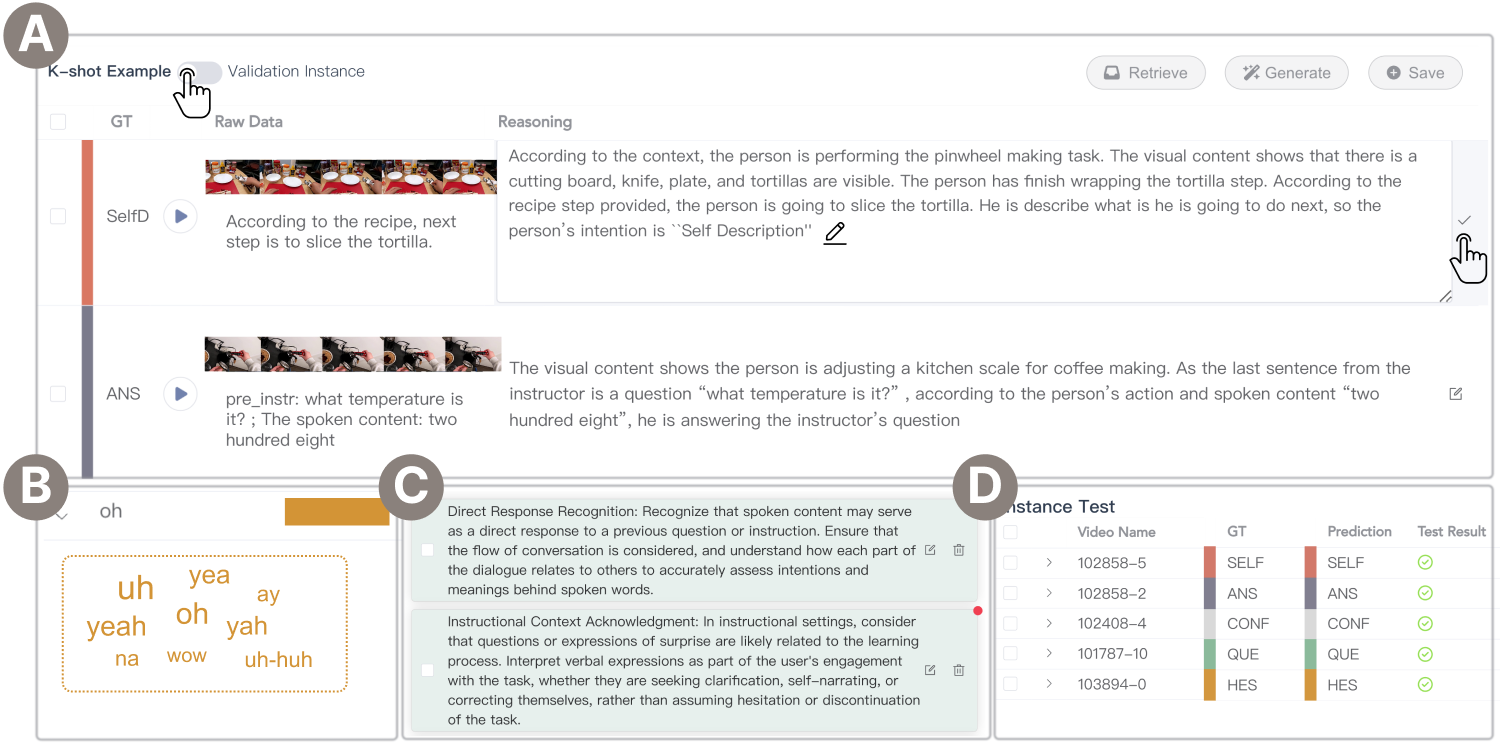}
    \caption{(A) The selected and annotated k-shot examples from distinct classes. (B) The ``uh'' pattern influenced the ``Hesitation'' class reasoning. (C) The recommended principles to guide ``Hesitation'' class reasoning. (D) The test results of added out-of-distribution instances.}
    \label{fig: casetwo}
\end{figure}

To address this problem, \textbf{E6} decided to include more explicit \jianben{rationales} of each prediction class within the prompt instructions \jianben{to guide the model}. (\textbf{R3}).
Thus, he revised the prompt to add the clarification such as \textit{``Self Description refers to scenarios where the user narrates or explains what they are doing, intend to do, or their thought process regarding the task at hand.''}
While submitting this prompt for testing, \textbf{E6} also thought that, besides \jianben{providing} explicit \jianben{rationales}, he could also \jianben{include} concrete k-shot examples to help the model learn (\textbf{R3}). 
He navigated to the \textit{K-shot Example Mode} in the \textit{Reasoning Panel} and selected five K-shot examples, each representing a distinct class from the top recommended ones (\cref{fig: casetwo}A). 
\textbf{E6} also noticed that the rationales generated by the more advanced auxiliary LLM also contained errors for the ``Self Description'' class, indicating that this category might be challenging for LLMs to grasp and reason about, underscoring the need for providing additional guidance in the prompt.
Following the refinement of rationales for the k-shot examples, \textbf{E6} imported these annotated examples and submitted this prompt version for testing. 
\textbf{E6} then examined the updated test outcomes in the \textit{Evaluation Panel} (\textbf{R2}). 
The increased \jianben{performance statistics} proved that providing either explicit explanations or k-shot examples can help improve the LLM's reasoning performance (~\autoref{fig: case_two_first}).

\textbf{E6} further explored the performance specifics of the latest prompt version (enhanced with k-shot examples) in the \textit{Reasoning Panel} (\textbf{R1}). \jianben{He identified} a cluster of errors \jianben{in the first layer of the Sankey diagram associated with the predicted ``Hesitation'' class.}
The consistent yellow color of the language modality and the overall prediction suggested that language modality predominated the reasoning process, and all these instances were misclassified as the ``Hesitation'' class.
In the pattern table on the right, he identified a frequent language pattern, ``uh'', associated with a high error rate, with its bar chart fully colored in yellow  (\cref{fig: casetwo}B). 
\jianben{Upon expanding the row, he found it contained evidence like ``uh'' and ``oh'' that indicated ``Hesitation''.}
Therefore, \textbf{E6} clicked the row to examine the specific instances it included. He found that whenever the spoken content contained modal words like ``uh'' and ``oh'', the model interpreted these as indicators of unwillingness to continue, thereby predicting the user's intention as ``Hesitation'' without considering any other factors. 
Consequently, \textbf{E6} selected these instances for the auxiliary LLM to summarize principles for avoiding such error (\textbf{R3}).
He then refined and incorporated these principles (\cref{fig: casetwo}C) into the prompt and saved these instances in the \textit{Instance Test} view. 
Additionally, he added multiple instances from his project into the \textit{Instance Test} view to evaluate the prompt robustness (\textbf{R2}). 
The test results showed that the accuracy reached 77\%, with the added test instances correctly predicted (\cref{fig: casetwo}D). 
\textbf{E6} was satisfied with this result and planned to use the prompt for his project.

\subsection{Expert Interviews}
We further conducted semi-structured interviews with two academic researchers and one industry research scientist (\textbf{P1-P3}) to verify the effectiveness and usability of \name.
All participants had experience in prompt engineering and the training or adaption of multimodal LLMs for downstream tasks, while none had previously tried the \name before the interviews.
Each interview began with the research background introduction, followed by the system workflow and function demonstration with examples.
Experts were then invited to freely explore the system using real datasets, voicing their thoughts in a think-aloud manner. \jianben{We also collected feedback from \textbf{E5} and \textbf{E6} during case studies}.
The gathered feedback is summarized below:

\textbf{System workflow} 
All experts concurred that the workflow of \name is thoughtfully designed, significantly improving the efficiency of prompt iteration compared to their current practices, \jianben{which rely} solely on performance statistics for evaluating prompt effects and laborious manual experiments to search for better-performing prompts.
As \textbf{P2} noted, ``I think \name offers a more systematic and comprehensive way to analyze the model's complex reasoning behaviors.''
\textbf{P1} highlighted that the varied strategies and streamlined process provided by \name notably ``reduce the pain for prompt writing and testing '' which are challenging tasks for them.
\textbf{E5} commented that the recommended principles and K-shot examples ``serve as good starting points to bring new perspectives and inspire thoughts''.

\textbf{System designs and interactions}
All experts remarked that the visual and interaction design of \name is intuitive and easy to learn and use.
\textbf{P3} expressed particular favor for the \textit{Prompt History} design, which makes it effortless to track every detail of changes, ``as I usually get lost after several rounds of prompt iteration. Now I can start with any version at ease.''
\textbf{P1} valued the convenient \jianben{one-click} generate and import function, which saves tons of time in manually editing and formatting the prompts. 
\textbf{E5} appreciated the ability to examine and evaluate at the instance level with reference to raw data, stating, ``Since hallucinations can happen inevitably, having access to instance-specific details for validation significantly increased my trust for the system and confidence in the prompts I developed''.
Meanwhile, experts \textbf{E6} and \textbf{P2} mentioned that it took some time to understand and proficiently use the Sankey diagram, yet they acknowledged that the complexity of multimodal reasoning performance necessitates such a design.

\textbf{Suggestions for improvement}
\textbf{P1} proposed that the generated instance-specific principles can be visually linked to their originating instances to offer a more intuitive and comprehensible reference.
\textbf{P2} expressed a desire for a feature that allows the system to recommend instances based on users' high-level input criteria for further evaluation or demonstration. 
\textbf{E6} also thought it would be beneficial if the system could help summarize users' annotated rationales to identify potential ambiguities and conflicts. 
\textbf{P3} thought it would be interesting and useful to enable comparisons across multiple LLMs.
Besides, step-by-step guides are wanted during real-time exploration to reduce learning curve.


\section{Discussion}
In this section, we discuss the \name regarding knowledge alignment with principle, system generalizability, and scalability. We also pointed out current limitations and potential directions for future work.

\textbf{Human-AI knowledge alignment through principle}
Given the emerging prompting paradigm that allows users to interact with LLMs through natural language, there is a growing interest in harnessing explicitly stated principles for evaluating and guiding model performance in downstream applications. 
While previous studies have explored the assessment of models using human-input criteria~\cite{kim2023evallm} and the alignment of chatbot behaviors with user preferences through converting feedback into principles~\cite{petridis2023constitutionmaker}, our research pioneers the use of data-derived principles to direct and \jianben{improve} model multimodal reasoning performance. 
Drawing on the innate human capacity for both inductive and deductive reasoning, \name proposed an LLM-assisted module condensing both instance-specific and agnostic principles to encourage users to efficiently express and externalize their domain-specific knowledge and expertise for model steering.
Despite the exhibited great potential for eliciting desired knowledge, exploring how to design, manage, and apply principles more effectively across varied tasks and contexts remains a fertile area for research. 
As pointed out by prior works~\cite{petridis2023constitutionmaker, zhang2024context}, there is no one-size-fits-all principle granularity, as the effectiveness varies with task complexity, dataset diversity, and principle quality. In our work, we provide both specific and universal principles for balancing both uniqueness and generability. Identifying and crafting an effective set of principles with suitable granularity for different tasks remains an open question. Moreover, current users can only articulate principles in natural language where more diverse interactions (e.g., clicking in SAM~\cite{kirillov2023segment}) can be integrated to enable users to provide more nuanced and precise feedback.
Meanwhile, managing the accumulated principles is non-trivial due to conflict and forgetting issues. Users may also struggle to grasp the influence of varying principles on model performance. Utilizing LLMs to condense and differentiate the patterns and impacts of principles could serve as a potential solution.
On the other side, while principles are most effective for large models possessing robust instruction-following capacities, they can also benefit smaller models by guiding dataset retrieval and generation for model fine-tuning. 
Furthermore, as demonstration examples and principles represent two distinct approaches of injecting and eliciting knowledge for reasoning in a bottom-up and top-down manner respectively, how to collocate k-shot examples with principles to maximize information gain in prompt engineering remains a compelling question.

\textbf{System generalizability and scalability}
In this paper, we mainly focus on the interaction between the two most-studied visual and language modalities. However, our system can be extended to investigate the interactions between multiple modalities by pair-wise comparison. 
Besides the analysis tasks evaluated in this paper, the proposed framework is readily to be utilized for other multimodal content comprehension and reasoning tasks such as multimodal hate or sarcasm recognition, and multimodal context question answering~\cite{yang2023mmbigbench}, where the modality interaction relationship persistently exist.
This prompting-based system can also serve as a testing tool to uncover weaknesses in model multimodal reasoning performance and identify example types and principles to inform larger-scale data collection for model fine-tuning.
Moreover, the design of the system can be extended for other applications. For example, the \textit{Reasoning Panel} design can be used for other tasks that necessitate summarizing relationships across various information channels at multiple levels.
The highlighted difference design in \textit{Prompt History} view can also help text summarization and comparison tasks in a structured and intuitive way.
The system scalability is rooted in the algorithm and visual design. The bottleneck of the algorithm part is the time cost of processing the video dataset and LLM's generation speed. Currently, we have implemented batch processing to expedite the data processing and generation process for a smooth prompting experience. However, this approach may not suffice for handling the data scale of thousands of instances, necessitating the exploration of strategies like parallel computing and data sampling to ensure instant feedback. 
For the visual design, The Sankey diagram design in \textit{Reasoning Panel} may become visually cluttered when dealing with a large number of prediction classes or complex modalities. For this situation, we can consider adopting a hierarchical visualization design coupled with interaction techniques to enhance visual scalability.

\textbf{Limitations \& Future Work}
Current LLMs exhibit deficiencies in producing hallucinated and inconsistent responses. Our system has tried to mitigate this issue by fixing hyperparameters and providing a multi-level systematic analysis of outputs regarding different prompts, allowing users to easily examine and identify outlying responses. Future efforts can be directed towards developing techniques for reducing hallucination occurrence in model outputs. 
Moreover, considering the potential information loss or inaccuracies introduced by expert models across different modalities, we plan to integrate more advanced expert models and visualize potential uncertainties to increase user trust. 
In the future, we consider enabling comparison across multiple LLMs to further investigate effective prompt engineering strategies for different models and tasks. Additionally, we plan to extend our work to study interaction involving more modalities in increasingly complex scenarios and applications.

\section{Conclusion}
In this paper, we introduce \name, a novel visual analytics tool designed to facilitate prompt engineering for enhancing multimodal reasoning of LLMs with human insight and expertise. The system allows users to thoroughly assess prompt effectiveness through well-summarized multimodal reasoning patterns and offers varied strategies for prompt revision, enabling users to apply their knowledge for efficient prompt iteration. The system's efficacy and efficiency are validated through two case studies and positive feedback from experts.

\bibliographystyle{abbrv-doi-hyperref}

\bibliography{main}








\end{document}